\documentclass[article,reprint,amsmath,longbibliography,amssymb,amsfonts,superscriptaddress,aps,prl]{revtex4-2}
\usepackage{times,graphicx,caption,epstopdf,dcolumn,bm,color,braket,multirow,xfrac,etoolbox}
\usepackage[colorlinks=true, allcolors=blue]{hyperref}
\usepackage[newcommands]{ragged2e}
\usepackage{amsmath}

\begin{document}

\title{Signatures of Amorphous Shiba State in FeTe$_{0.55}$Se$_{0.45}$}

\author{Jinwon Lee}
\altaffiliation{\texorpdfstring{These authors contributed equally to this work}{}}
\affiliation{Leiden Institute of Physics, Leiden University, Leiden 2333CA, The Netherlands}

\author{Sanghun Lee}
\altaffiliation{\texorpdfstring{These authors contributed equally to this work}{}}
\affiliation{Department of Physics, Yonsei University, Seoul 03722, Republic of Korea}

\author{Andreas Kreisel}
\altaffiliation{\texorpdfstring{These authors contributed equally to this work}{}}
\affiliation{Center for Quantum Devices, Niels Bohr Institute, University of Copenhagen, Copenhagen {\O} 2100, Denmark}

\author{Jens Paaske}
\affiliation{Center for Quantum Devices, Niels Bohr Institute, University of Copenhagen, Copenhagen {\O} 2100, Denmark}

\author{Brian M. Andersen}
\affiliation{Center for Quantum Devices, Niels Bohr Institute, University of Copenhagen, Copenhagen {\O} 2100, Denmark}

\author{Koen M. Bastiaans}
\affiliation{Leiden Institute of Physics, Leiden University, Leiden 2333CA, The Netherlands}

\author{Damianos Chatzopoulos}
\affiliation{Leiden Institute of Physics, Leiden University, Leiden 2333CA, The Netherlands}

\author{Genda Gu}
\affiliation{Condensed Matter Physics and Materials Science Department, Brookhaven National Laboratory, Upton, NY 11973, USA}

\author{Doohee Cho} \email{dooheecho@yonsei.ac.kr}
\affiliation{Department of Physics, Yonsei University, Seoul 03722, Republic of Korea}

\author{Milan P. Allan} \email{allan@physics.leidenuniv.nl}
\affiliation{Leiden Institute of Physics, Leiden University, Leiden 2333CA, The Netherlands}
\affiliation{Faculty of Physics, Ludwig-Maximilians-University Munich, Munich 80799, Germany}
\affiliation{Center for Nano Science (CeNS), Ludwig-Maximilians-University Munich, Munich 80799, Germany}
\affiliation{Munich Center for Quantum Science and Technology (MCQST), Ludwig-Maximilians-University Munich, Munich 80799, Germany}

\begin{abstract}
The iron-based superconductor FeTe$_{0.55}$Se$_{0.45}$ is a peculiar material: it hosts a surface state with a Dirac dispersion, is a putative topological superconductor hosting Majorana modes in vortices, and has an unusually low Fermi energy. The superconducting state is generally thought to be characterized by three gaps in different bands, with the usual homogenous, spatially extended Bogoliubov excitations -- in this work, we uncover evidence that it is instead of a very different nature. Our scanning tunneling spectroscopy data shows several peaks in the density of states above a full gap, and by analyzing the spatial and junction-resistance dependence of the peaks, we conclude that the peaks above the first one are not coherence peaks from different bands. Instead, comparisons with our simulations indicate that they originate from generalized Shiba states that are spatially overlapping. This can lead to an amorphous state of Bogoliubov quasiparticles, reminiscent of impurity bands in semiconductors. We discuss the origin and implications of this new state.
\end{abstract}

\maketitle

\begin{figure}[tb]
\includegraphics[width=\linewidth]{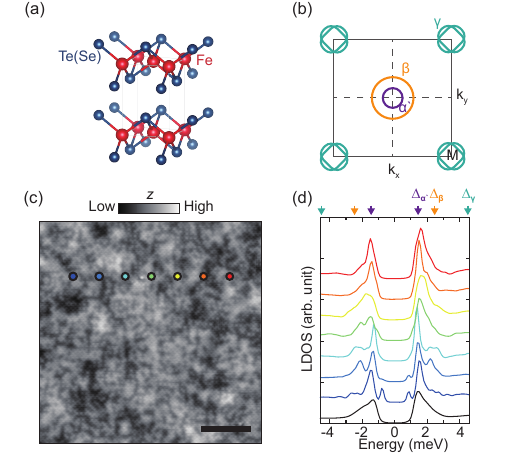}
\caption{The disordered superconductor FeTe$_{0.55}$Se$_{0.45}$. (a)~Schematic of the FeTe$_{0.55}$Se$_{0.45}$ crystal structure.
(b)~The Fermi contours of FeTe$_{0.55}$Se$_{0.45}$. Two hole-like bands ($\alpha$' and $\beta$) at the center of the Brillouin zone and the folded electron-like band ($\gamma$) at the corners of the Brillouin zone.
(c)~The STM topography of FeTe$_{0.55}$Se$_{0.45}$ taken with $V_{\rm set}$ = --6~mV and $I_{\rm set}$ = 120~pA. Scale bar, 5~nm.
(d)~The differential conductance ($\mathrm{d}I/\mathrm{d}V$) measured at the sites marked by colored dots in (c).
 The black curve is a spatially averaged spectrum. Each curve is shifted for clarity. The colored arrows indicate the multiple superconducting gaps assigned by previous ARPES measurements~\cite{zhang2019multiple}.}
\label{Fig.1}
\end{figure}

FeTe$_{0.55}$Se$_{0.45}$ exhibits a number of unusual properties in the superconducting state~\cite{wang2011electron,hirschfeld2011gap,hanaguri2010unconventional,wang2015topological,rinott2017tuning,zhang2018observation,zhang2019multiple}. It is highly disordered, basically being an alloy of FeSe and FeTe [Fig.~\ref{Fig.1}(a)], and it remains to be resolved exactly how much remnant magnetism exists. Despite this, superconductivity is robust, with a transition temperature $T_{\rm c}$ of 14.5 K. The pairing symmetry is widely believed to be s$_\pm$~\cite{hirschfeld2011gap,hanaguri2010unconventional,chen2019direct} with different gap sizes on different bands/orbitals. There are two hole-pockets at the $\Gamma$ point and two electron pockets at the M point; the smaller (larger) hole pocket is referred to as $\alpha$' ($\beta$), and the electron pocket $\gamma$ band [Fig.~\ref{Fig.1}(b)]. Recent angle-resolved photoemission spectroscopy (ARPES) studies revealed that the $\alpha$', $\beta$, and $\gamma$ bands have a superconducting gap with energy values of roughly 1.4, 2.4, and 4.5~meV, respectively~\cite{miao2012isotropic}. One would expect that scanning tunneling microscopy (STM) and spectroscopy reflect the multiple-gap structure determined by several coherence peaks similar to other multi-band superconductors like LiFeAs~\cite{li2022ordered} or NbSe$_{2}$~\cite{noat2015quasiparticle}: two sets of sharp coherence peaks with the larger gapping out a proportion of the spectral weight, and the smaller one the rest. Previous STM data on FeTe$_{0.55}$Se$_{0.45}$, taken with less energy resolution, appeared to show just one single gap of 1.7~meV~\cite{hanaguri2010unconventional}. However, better energy resolution revealed a much more complex set of peaks, reminiscent of – and previously identified as – coherence peaks of the Bogoliubov density of states (DOS) from the different bands~\cite{machida2019zero,wang2020evidence}. It was conjectured that the significant spatial variations of the spectral features are due to disorder causing electronic heterogeneity~\cite{singh2013spatial}.

In this Letter, we demonstrate that our high-energy-resolution spectroscopic data is inconsistent with an interpretation of multiple coherence peaks. The local density of states (LDOS) mapping reveals that the observed peaks are unlike a single coherence peak but spatially dispersive and influenced by the electric field of the tip, similar to low-energy Shiba states~\cite{chatzopoulos2021spatially}. These experimental observations let us exclude the possibility that the spectral peaks beyond the first ones are the coherence peaks of the larger superconducting gaps. Instead, based on good agreement with simulations, we argue that these peaks are generalized Shiba states~\cite{shiba} associated with a larger gap. 
They are randomly distributed, and migth be caused by the random distribution of the two different chalcogen atoms~\cite{singh2013spatial,wang2021surface,zhao2021nematic}, or the remnant non-periodic spin structure in the iron lattice~\cite{aluru2019atomic}. Either can produce a landscape with emergent locations favorable for generating generalized Shiba states. These states have substantial spatial overlap and are likely to form an amorphous quasiparticle state below the largest superconducting gap. The ramification of this scenario is that the low-energy electronic properties of FeTe$_{0.55}$Se$_{0.45}$ are characterized by an amorphous Shiba state. 

\captionsetup{justification=Justified}

\captionsetup{justification=Justified}  
\begin{figure*}[tb]
\includegraphics[width=\linewidth]{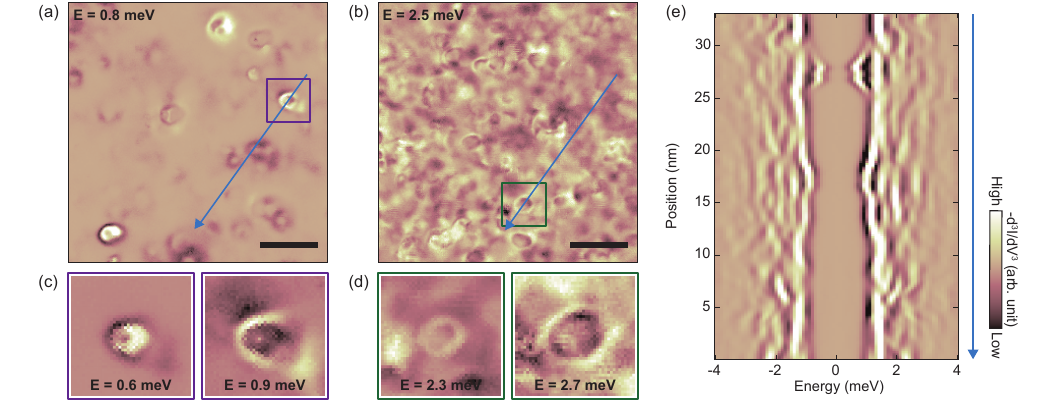}
\caption{Spatially dispersive generalized Shiba states inside and outside the superconducting gap. (a) and (b) Negative second derivative of deconvoluted differential conductance ($-\mathrm{d}^{3}I/\mathrm{d}V^{3}$) maps at the energies of +0.8 and +2.5~meV for the same field of view. The $\mathrm{d}I/\mathrm{d}V$ maps are acquired with $V_{\rm set}=-6$~mV, $I_{\rm set}=1.20$~nA, and $V_{mod}=71$~$\mu$$V_{\rm rms}$. Scale bar, 10 nm. (c) and (d) Spatial evolution of the ring patterns as a function of energy, corresponding to the purple and green box in (a) and (b) respectively. (e) The $-\mathrm{d}^{3}I/\mathrm{d}V^{3}$ spectra taken along the blue arrow in (a) and (b) (each line averaged over $3\times3$ pixel in the map).}
\label{Fig.2}  
\end{figure*}

\captionsetup{justification=Justified} 
\begin{figure}[tb]
\includegraphics[width=\linewidth]{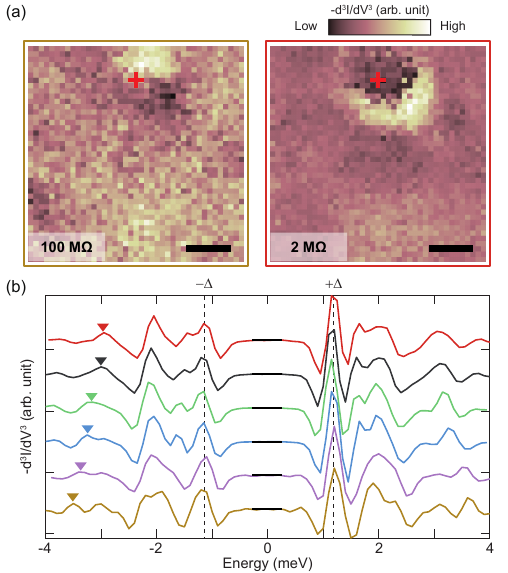}
\caption{$R_{\rm J}$-dependent energy dispersion of the out-gap state. (a)~$R_{\rm J}$-dependent spatial extent of the out-gap impurity state at the energy of +3.5~mV. Scale bar, 1~nm. (b)~Point $-\mathrm{d}^{3}I/\mathrm{d}V^{3}$ spectra obtained at the location marked with a red cross in (a) with decreasing $R_{\rm J}$ [$R_{\rm J}=2$ (top), 2.5, 5, 10, 50, and 100~M$\Omega$ (bottom)]. Each curve is spatially averaged within $3\times3$ pixels, then shifted for clarity (zero values are indicated by black solid lines).}
\label{Fig.3}  
\end{figure}

To arrive at these findings, we cleave FeTe$_{0.55}$Se$_{0.45}$ crystals~\cite{wen2011interplay} in an ultra-high vacuum at about 30 K and perform spectroscopic-imaging STM experiments at 2.2 K. For increased energy resolution, we use a tip with a superconducting apex with a gap of $\Delta_{\rm tip} = 1.3$~meV and a broadening of $\Gamma=45$~$\mu$eV~\cite{cho2019strongly}.
The spectra shown in Figs.~\ref{Fig.1}-\ref{Fig.3} are deconvoluted using the tip spectrum obtained from experiments on a Pb(111) surface as described in Ref.~\cite{chatzopoulos2021spatially} (see  Supplementary Figs.~1-3~\cite{suppl} for non-deconvoluted spectra).
To understand the observations theoretically, we start from a multi-band model of the electronic structure, include a superconducting interband and intraband order parameter as expected from spin-fluctuation pairing and calculate the LDOS by including disorder within a T-matrix approach. See Supplemental Material for more details~\cite{suppl}.

Figure~\ref{Fig.1}(c) shows the topography of a cleaved surface with a square lattice composed of chalcogen atoms. The absence of protrusions in the STM image and the high $T_{\rm c}$ of our samples indicates that our sample includes very little excess Fe atoms~\cite{hanaguri2010unconventional,kato2009local,yin2015observation} – which suppresses $T_{\rm c}$~\cite{wen2011interplay} – consistent with previous STM studies~\cite{machida2019zero,wang2020evidence,singh2013spatial}. Figure~\ref{Fig.1}(d) shows a few spectra acquired at the sites marked with colored dots in Fig.~\ref{Fig.1}(c) in agreement with previous work~\cite{hanaguri2010unconventional,wang2020evidence,singh2013spatial}. While the average spectrum [black curve in Fig.~\ref{Fig.1}(d)] looks somewhat similar to the Bogoliubov DOS expected from a single-gap Bardeen-Cooper-Schrieffer superconductor, our high-energy-resolution spectroscopy reveals that there is more than one peak in most single spectra [colored curves in Fig.~\ref{Fig.1}(d)]. As the number of peaks and their energy change with location, the average spectrum will appear as if there is a single coherence peak with a broad hump (or shoulder). While these spectral features have previously been attributed to multiple superconducting gaps at different bands~\cite{wang2020evidence}, we show here that our tunneling spectra are neither a reflection of a single gap nor of a series of coherence peaks. Instead, we argue that these spectral peaks stem from generalized Shiba states.

The first argument against the interpretation that the peaks manifest multiple coherence peaks at different bands stems from the ring-like features [Fig.~\ref{Fig.2}(a)] for the negative second derivative of the $\mathrm{d}I/\mathrm{d}V$ spectra ($-\mathrm{d}^{3}I/\mathrm{d}V^{3}$, see Supplementary Figs. 4 and 5 for the same plots with $\mathrm{d}I/\mathrm{d}V$ spectra~\cite{suppl}). This is reminiscent of the generalized Shiba states observed previously at much lower energies of 0.8~meV (inside the lowest energy peak) and at much lower densities~\cite{chatzopoulos2021spatially}. There, similar ring-like features can be explained by impurity levels influenced by tip-gating. Looking at the peaks outside $\Delta_{1}$, we observe similar ring-like features [Fig.~\ref{Fig.2}(b)]. They are overlapping due to their high density.
Figures~\ref{Fig.2}(c)-(e) reveal that the ring-like features are attributed to spatially dispersive spectral features with particle-hole symmetry (Supplementary Fig.~6~\cite{suppl}); that is, their sizes vary depending on energy.  While this further suggests their association with a superconducting state, their spatially dispersive behaviors contradict expectations of superconducting coherence peaks. Instead, they point to similar physics as seen in previously reported in-gap Shiba states~\cite{chatzopoulos2021spatially}, while originating from a much denser set of impurities. We further note that the absence of the spatial correlation between in-gap and out-gap Shiba states indicates that they have different origins.

The second argument stems from the junction resistance ($R_{\rm J}$) dependence. We compare the $-\mathrm{d}^{3}I/\mathrm{d}V^{3}$ maps at the energy of +3.5 meV obtained with vastly different $R_{\rm J}$ in Fig.~\ref{Fig.3}(a).  We obtain all these data with constant-current mode to sustain the same $R_{\rm J}$ for long-term measurement while keeping the same set bias ($V_{\rm set}$) and increasing (decreasing) the set current ($I_{\rm set}$) to decrease (increase) the $R_{\rm J}$. It is noticeable that the ring-like patterns change depending on the $R_{\rm J}$; it becomes larger (smaller) when the $R_{\rm J}$ is smaller (larger). This is because the energy level of the impurity state is tuned by the $R_{\rm J}$, as shown in Fig.~\ref{Fig.3}(b). If the tip is brought closer to the sample (lower $R_{\rm J}$), a larger electric field is applied at the surface with the same $V_{\rm set}$. This is not surprising because the potential caused by the electric field from the STM tip is expected to be a function of the distance between the tip and the impurity, and it can vary as we move the tip either in a lateral or vertical direction~\cite{chatzopoulos2021spatially,farinacci2018tuning}. Thus, we have a similar trend in the size of the ring-like feature when we look at different energies with the fixed $R_{\rm J}$ and when we look at different $R_{\rm J}$ with the fixed $V_{\rm set}$. We emphasize that while the tip “gates” the generalized Shiba states, allowing us to distinguish them from coherence peaks, their presence is independent of the tip.

The third argument stems from the spectral shapes and their spatial variations. In the absence of impurities, we would expect that each gap decreased the DOS further, leading to the highest intensity of the outermost peak, as observed in the multi-band superconductors LiFeAs~\cite{li2022ordered} or NbSe$_{2}$~\cite{noat2015quasiparticle}. However, this is not the case for most ($\sim90$~$\%$) of our spectra showing the highest intensity at the smallest superconducting gap. This could be explained by tunneling matrix elements determined by constituent orbitals of the multi-bands~\cite{takahashi2016orbital,kim2017atomic,yin2020orbital}. This scenario would, however, require that the large gap is on the $d_{\rm xy}$ orbital, which is in-plane and exhibits smaller tunneling~\cite{kreisel2021quasi}, an unexpected situation for iron-based superconductors where the large gap is usually on the $d_{\rm xz}/d_{\rm yz}$ orbitals. Additionally, in the strained surface of LiFeAs, the $\mathrm{d}I/\mathrm{d}V$ spectra also exhibit enhanced spectral weight at the smallest superconducting gap and suppression of the outermost peak~\cite{li2022ordered}. This suggests a possible link between the inhomogeneity caused by local strain from different configurations of chalcogen atoms~\cite{singh2013spatial} and the observed suppression of the outermost peak, which we discuss below. Moreover, the fact that the number of peaks varies across locations disagrees with the expectation from a multi-band superconductor.

Taken together, our data is inconsistent with the hypothesis of the extra peaks in the DOS of FeTe$_{0.55}$Se$_{0.45}$ being the coherence peaks of a multi-band superconductor. We further mention the inconsistency with another possibility: inelastic tunneling due to bosonic modes, which could lead to the appearance of extra peaks. In such cases, the additional peaks follow the sharp coherence peak with a certain energy spacing corresponding to the energy scale of the bosonic modes, e.g. phonons~\cite{schackert2015local,jandke2016coupling} or spin-fluctuations~\cite{chi2017imaging,hlobil2017tracing}, involved in pairing electrons. However, except for a few positions, there is no fixed energy between the peaks. Furthermore, such peaks from inelastic tunneling would likely be visible above the critical temperature as well~\cite{jandke2016scanning}, in disagreement with previous STM measurements on FeTe$_{0.55}$Se$_{0.45}$~\cite{hanaguri2010unconventional}.

Instead, we argue here that the phenomenology in FeTe$_{0.55}$Se$_{0.45}$ points towards an amorphous Shiba state causing a spectral weight transfer from the coherence peak of the larger superconducting gap to a set of Shiba states within this gap but larger than the smallest superconducting gap. These states overlap -- we discuss the possibly important consequences of this fact at the end of our Letter.
To consider such a scenario with the peaks being generalized Shiba states, we now have to address (i) why the coherence peak of the larger gap is not visible, (ii), how impurity-bound states can exist outside of the smallest superconducting gap ($\Delta_{1}$), and (iii) whether realistic parameters for impurity states can lead to extended peaks at energies $E>\Delta_{1}$. To answer these questions, we turn to theoretical simulations.


\captionsetup{justification=Justified}
\begin{figure}[tb]
\includegraphics[width=\linewidth]{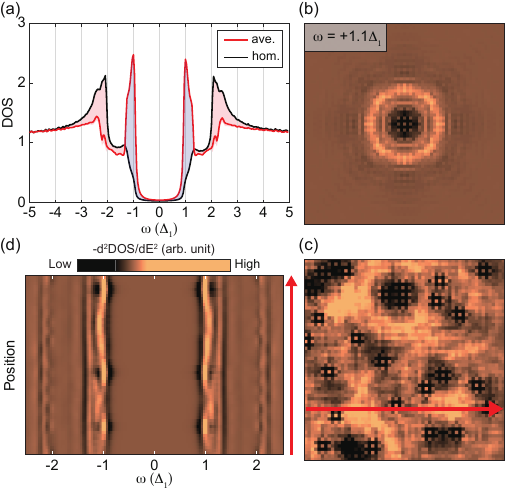}
\caption{Orbital-selective pair-breaking in a multi-band Fe-based superconductor. (a)~DOS of the homogeneous system compared to the average DOS of the inhomogeneous system where the coherence peak at $\Delta_{1}$ is enhanced (blue shaded) while the coherence peak at $\Delta_{2}$ is suppressed (red shaded). (b) and (c) The negative second derivative of LDOS maps on a field of view of $60\times60$ lattice constants at the energy $\omega=1.1\Delta_{1}$, showing a single impurity and 30 impurities, respectively. (d)~The negative second derivative of the LDOS along the cut marked by the red arrow in (c) showing inhomogeneous peaks and the spatial dispersion from the generalized Shiba states at $\omega>\Delta_{1}$ (each line averaged over $3\times3$ pixel in the map)}
\label{Fig.4}
\end{figure}

For our simulation of the strongly disordered FeTe$_{0.55}$Se$_{0.45}$ system, we use a generic five-band model
with three orbitals, $d_{\rm xy}$, $d_{\rm xz}$, and $d_{\rm yz}$, present at low energies. Employing an interorbital and intra-orbital order parameter, two energy scales are associated with superconductivity. In the order parameter of s$_{\pm}$ symmetry two gap scales $\Delta_{1}$ and $\Delta_{2}$ occur, similar as found in photoemission experiments~\cite{miao2012isotropic}. We treat the impurities in the single-impurity approximation and include two different types of nonmagnetic scatterers (note that magnetic scatters would lead to a similar phenomenology). To this end, we set up a \textit{T}-matrix calculation of the LDOS where the local impurity potential is influenced by the presence of the STM tip by the induced gating potential $V_{\rm imp}({\mathbf r})$, where ${\mathbf r}$ is the distance of the STM tip to the impurity. The DOS picked up by the STM tip is calculated for a single impurity, and the density modulations are superimposed for a representative distribution of impurities (see  Supplemental Material for details~\cite{suppl}). The homogeneous DOS features two coherence peaks close to $\Delta_{1}$ and $\Delta_{2}$ [Fig.~\ref{Fig.4}(a)], with the larger gap removing around half of the spectral weight. This is reminiscent of clean multi-band superconductors such as LiFeAs~\cite{li2022ordered}, but very different from our observations in FeTe$_{0.55}$Se$_{0.45}$. Instead, as discussed below, the strongly disordered network of overlapping Shiba states alters the DOS to a single-gap DOS with only one visible peak at $\Delta_{1}$.

Figure~\ref{Fig.4} summarizes the outcome of the theoretical simulations of the strongly disordered FeTe$_{0.55}$Se$_{0.45}$ superconductor. Figure~\ref{Fig.4}(a) compares the spatially averaged DOS, whereas Figs.~\ref{Fig.4}(b) and \ref{Fig.4}(c) display the LDOS maps for a single impurity and multiple ones at energies between $\Delta_{1}$ and $\Delta_{2}$. Finally, in Fig.~\ref{Fig.4}(d), we show representative cross-sections. Additional cross-sections of other impurity configurations are provided in the Supplemental Material~\cite{suppl}, where we also discuss the properties of single impurities of two different types. 

The simulations for multiple impurities display a striking resemblance to our data in several key aspects. First, the non-magnetic impurities produce in-gap states, as also observed experimentally. Second, due to the gating effect of the STM tip, the states disperse as a function of tip position and therefore extend as rings of several nanometers around the impurities if examined as conductance maps, see Fig.~\ref{Fig.4}(b). Third, and perhaps most surprisingly, the superposition of impurity-bound states [Fig.~\ref{Fig.4}(c)] and associated peaks and dips in the spectrum leads to a strong modification of the homogeneous DOS as shown in Fig.~\ref{Fig.4}(a). The spectral weight of the coherence peak at $\Delta_{2}$ is drained (red shaded area), leaving only a faint trace; while the peak close to $\Delta_{1}$ is strongly enhanced (blue shaded area) rendering it similar in overall appearance to the spectrum of a single-gap superconductor, at least when viewed from the spatially averaged spectrum [Fig.~\ref{Fig.1}(e)]. All these features are seen in the experimental data as well. For a more detailed comparison, we also provide spectra along cuts in real space where it is visible that the impurity-bound states of the strong impurities disperse at energies below $\Delta_{1}$ (see Supplemental Material~\cite{suppl}), while there are dispersing peaks beyond the energy scale $\Delta_{1}$ stemming from the weak type of impurities. 

Note that our modeling uses a large impurity potential for one type of scatterers, diagonal in orbital space which could be associated with defects on the Fe site (also below the surface layer) and will therefore act as a strong scatterer. The other weak impurity is chosen to only affect the $d_{\rm xz}$ and $d_{\rm yz}$ orbitals in our model because the orbitals can be coupled to impurities away from the Fe plane. This yields the phenomenology of not inducing any bound states at energies below $\Delta_{1}$ (as in the experimental data). This choice is compatible with the effects of the disorder at Se/Te sites. Disorder in the chalcogen layer mainly affects the inter-layer coupling of the $d_{\rm xz}$ and $d_{\rm yz}$ orbitals since these hybridize with the Se/Te atoms out of the plane. In contrast, the $d_{\rm xy}$ orbital of Fe couples mostly to the neighboring Fe atoms and less in the out-of-plane direction. Thus, we propose that the scattering is orbital-selective, dominantly couples to the states on the Fermi surface where the order parameter is large, i.e., beyond the scale $\Delta_{1}$, and therefore effectively suppresses signatures from the larger order parameter $\Delta_{2}$. This picture indicates that the disorder created by the FeSe/FeTe alloying could play a role. To simulate this, one would need a full real-space calculation in the field of view, which is out of reach with our current calculation power. At this point, we are agnostic to the origin of the scatterers, and note that Te/Se disorder or non-periodic spin structures on the iron atoms could play a role. The latter is reminiscent of experiments with magnetic islands on a conventional superconductor Nb(110) surface~\cite{goedecke2022correlation}. Taken together, the similarity between the simulations and data in all aspects is strong.

To conclude our paper, we would like to discuss the ramifications of a set of Shiba states to the physics of FeTe$_{0.55}$Se$_{0.45}$. The key aspect is that the states we detect are randomly distributed and have a strong spatial overlap. If these states are hybridized, this would lead to an amorphous Shiba band, not in the sense of a Bloch band but similar to the impurity band in semiconductors.
It would be without periodicity, embedded in a crystal lattice, and coexisting with electronic Bloch bands from said lattice. The interconnected impurities result in an unusual Hamiltonian, potentially leading to topologically nontrivial states in amorphous matter~\cite{franca2024topological,poyhonen2018amorphous,corbae2023amorphous}.
We do observe a few cases where two isolated but nearby Shiba states seem to be hybridized (Supplementary Fig.~7~\cite{suppl}), but generally, the high density of Shiba states and their complexity in energy and position obscure such signatures.
It would also be highly interesting to re-visit this issue by engineering amorphous networks on low-dimensional metal surfaces by atomic manipulations~\cite{nadj2014observation,kezilebieke2020topological,schneider2021topological,schneider2022precursors,Liebhaber2022,soldini2023two} and compare the results to FeTe$_{0.55}$Se$_{0.45}$. In any case, our work implies that the low-energy quasiparticle continuum of FeTe$_{0.55}$Se$_{0.45}$ is very distinct from that arising from a basic hopping-on-a-lattice model relevant to homogeneous multi-gap superconductors, and instead represents an intriguing new configuration of electron matter.

\begin{acknowledgments}
S. L. and D. Cho were supported by the National Research Foundation of Korea (NRF) grant funded by the Korea government (MSIT) (No. RS-2023-00251265 and RS-2024-00337267)
A.K. acknowledges support by the Danish National Committee for Research Infrastructure (NUFI) through the ESS-Lighthouse Q-MAT.
\end{acknowledgments}


%

\end{document}